# Towards a Grid Platform for Scientific Workflows Management

**Alexandru Costan*, Corina Stratan***
**Eliana Tîrşa*, Mugurel Ionut Andreica*, Valentin Cristea***

*University Politehnica of Bucharest, Splaiul Independenţei 313, Bucharest, Romania*
e-mail: {alexandru.costan,corina.stratan,eliana.tirsa,mugurel.andreica,valentin.cristea}@ cs.pub.ro

**Abstract:** Workflow management systems allow the users to develop complex applications at a higher level, by orchestrating functional components without handling the implementation details. Although a wide range of workflow engines are developed in enterprise environments, the open source engines available for scientific applications lack some functionalities or are too difficult to use for non-specialists. Our purpose is to develop a workflow management platform for distributed systems, that will provide features like an intuitive way to describe workflows, efficient data handling mechanisms and flexible fault tolerance support. We introduce here an architectural model for the workflow platform, based on the ActiveBPEL workflow engine, which we propose to augment with an additional set of components.

## 1. INTRODUCTION

Distributed applications, both in the academic and enterprise environments, are becoming more and more complex, requiring the orchestration of multiple services or programs into workflows. Workflow systems are built in order to assist the user in developing complex applications at a higher level, by organizing the components and specifying the dependencies among them.

Nowadays, commercial workflow engines provide a wide range of features suitable for enterprise applications. For scientific applications, even though a number of open source workflow systems are available, many of them are too difficult to use for non-specialists (some of them lack a graphical interface), or are restricted to a specific type of applications or on a single middleware platform; these problems have been impeding the adoption of workflow-based solutions in the scientific community.

Our purpose is to develop a workflow management platform for distributed systems, targeted to scientific applications, that will provide solutions for the following aspects:

- an intuitive way to describe workflows, based on ontologies specific to the application domains, allowing the user to work with abstract components

- flexible workflow structure, allowing the orchestration of services and also of plain executable programs efficient mechanisms for data handling, as scientific applications

- usually produce significant amounts of data; the mechanisms will be based on the data replication services provided by the underlying middleware

- comprehensive fault tolerance support, with configurable policies; as semantics and side effects vary from one application to another, we believe that the users should be able to select from multiple fault tolerance approaches the one that is the most suitable for a particular workflow

The workflow management platform will have three main components: a high-level module that will provide a user interface for defining abstract workflow, and that will manage domain-specific ontologies; a middle-level module that will have the role of a workflow engine; and a low-level module that will be in charge of scheduling the workflow activities and services onto the distributed system's physical resources, relying upon the available middleware.

Our focus in this work is on the middle-level module, the workflow engine. We have started by studying the facilities offered by the most commonly used workflow engines for scientific applications, from the point of view of the requirements presented above. Although some workflow engines provide advanced features for abstract workflows, data management or fault tolerance, they lack functionality in what concerns the other aspects. As a consequence, we consider the approach of starting from an existing open source workflow engine and implementing additional functions that are required for the purposes of our project. The engine we have studied is ActiveBPEL, one of the most widely used engines for WS-BPEL. We introduce here an architectural model of the modified ActiveBPEL engine, augmented with a new set of modules that will implement the additional functions.

The following part of this paper, Section 2, presents the related work in the field of workflow engines using Grids. Section 3 introduces the state-of-the-art data management functionalities in the existing workflow engines. In Section 4 we present the most commonly used fault tolerance mechanisms and our approach for this aspect. Section 5 introduces our proposed extensions to the ActiveBPEL architecture, and in Section 6 we summarize the conclusions of this study and the future work directions.

## 2. RELATED WORK

In this section we present a summary about interoperability between some of the workflow engines most used in scientific applications and middleware platforms. Condor DAGMan (Thain *et.al.*, 2005) submits jobs directly to the Condor scheduler; it doesn't offer support for other middleware.

Karajan (Laszewski *et.al.*, 2005) provides interoperability through the use of "providers" that allow middleware selection at runtime: GT2, GT3, GT4 or Condor. It has also support for SSH protocol. Authentication is done with either user certificates (personal mode) or host certificates (shared mode).

In Taverna (Hull *et.al.*, 2006) and ActiveBPEL workflow are seen as web services. The difficulty of implementation is hidden, users are presented a high-level interface. Interoperability for Taverna is limited to MyGrid, while ActiveBPEL can submit jobs to any middleware offering web services.

Triana (Taylor *et.al.*, 2006) is middleware agnostic; supports P2P, web services and Grids. GridLab GAT (Grid Application Toolkit), Triana's API for accessing Grid services, is written in such a way that new modules can be added, to achieve interoperability with different middleware platforms. Triana jobs do not have web interfaces, communication is done only through the input/output files, and submission is performed by a resource manager (GRAM1 or GRMS2). Triana can generate files entry for Pegasus / Condor.

Pegasus (Singh *et. al.*, 2005) sends its workflow to Condor DAGMAN / CondorG, in order to submit remote jobs. Pegasus users don't access DAGMan directly, except for optimization and troubleshooting.

Swift (Zhao *et.al.*, 2007) uses Globus Toolkit to submit jobs in Grid. For authentication and authorization on remote sites, it uses Grid Security Infrastructure (GSI).

P-GRADE Grid portal (Kertesz *et.al,* 2006) hides the details of low-level access to Grid resources, offering an interface which can be used with Globus Toolkit 2, Globus Toolkit 4, LCG-2 and gLite. Access to various Grids can also be done simultaneously, if the user certificates for those Grids are valid.

Many workflow engines work over a single type of middleware, besides those that enable web service orchestration (using WS-BPEL, for example) and should work with any middleware providing web services. This is one of the main reasons for choosing WS-BPEL as the specification language for our platform. We therefore analysed from the functional point of view the existing workflow languages. We noticed that WS-BPEL and Karajan are the most complex languages supporting a large number of basic models. We chose WS-BPEL for our proposed engine since it is a standardized language that provides support for many features and it is very expressive.

## 3. DATA HANDLING IN SCIENTIFIC WORKFLOWS

The workflow lifecycle consists of multiple phases in which data has a central role: a workflow generation phase where the analysis is defined, the workflow planning phase where resources needed for execution are selected, the workflow execution part, where the actual computations take place, and the result, metadata, and provenance storing phase. During workflow creation, appropriate input data and workflow components need to be discovered. During workflow mapping and execution data need to be staged-in and staged-out of the computational resources. As data are produced, they need to be archived with enough metadata and provenance information so that they can be interpreted and shared among collaborators.

During workflow creation, scientists specify the applications or workflows they want to run and the input data sets for these computations using unique logical identifiers or metadata, independent of where these data sets or analysis codes may be physically located in the distributed environment. Discovery of data sets, application codes, workflow templates, etc., is often done by querying various catalogs. Metadata catalogs store attributes that describe the contents of data sets. Provenance catalogs (Miles *et. al.*, 2007a) store information about computations and workflows to provide a detailed record of how analyses are run, including information about inputs to computations, application parameters used, calibration values for equipment, versions of workflow and analysis software used, etc. A challenging aspect of setting up these discovery catalogs is the need for communities to agree on standards for specifying metadata and provenance. Efforts to facilitate this are underway (Miles *et. al.*, 2007b).

In the workflow planning stage, the logical identifiers for applications and data must be mapped to resources in the distributed environment. For data sets that are inputs to workflows or analysis, this requires discovering the location of one or more copies of the desired data sets, selecting among them, and often copying or staging the data sets onto resources where computations will run. A scheduler is responsible for selecting among available data sets, selecting appropriate computational resources to run each task of a workflow, and orchestrating the movement of data sets and the execution of workflow tasks. Schedulers or workflow mappers need to be able to optimize the workflows based on some user-specified criteria. A major challenge in todays applications is the physical management of data in the distributed environment. Typically, as mentioned, replica location or metadata catalogs record mappings from logical identifiers for data to one or more physical locations where copies of the data sets are stored. Based on knowledge of the state of resources (the latency, bandwidth and load of storage systems, network bandwidth among nodes, etc.) that may be provided by information services, the workflow planner selects among available data replicas. In particular, the planner may try to select copies of the data that are close to the computational resources where workflow tasks will run, with respect to network latency or other metrics. It may be advantageous for workflow planning and execution services to coordinate with data placement services, whose role is to move data asynchronously with respect to workflow execution with the goal of improving the execution time of workflows. For example, a workflow engine might provide hints to a data placement service about required data sets as well as the expected ordering of data set access, based on knowledge or dependencies in the workflow.

Based on these hints, the placement engine can asynchronously stage some of the data required by the workflow engine onto shared storage resources near where the workflow tasks will execute. Workflows rely on a variety of data transfer mechanism over the wide area. These include such tools as GridFTP, the Fast Data Transfer (FDT) service, and others. In order to support the data transfer needs of their users and load balance the requests, many grid installations deploy multiple data movement servers targeting the same storage system.

In the workflow execution stage, an execution manager (ex: DAGMan) keeps track of tasks that must run and the dependencies among them. Earlier tasks in the workflow may produce intermediate data products that are consumed by tasks that run later. These intermediate data products may need to be staged from the resource where the earlier task ran to the resource on which the later task will run. The workflow execution system delays execution of a particular task until all its input data products are available on the computational resources where the task will run.

We briefly present as follows a few popular workflow engines from the point of view of their data handling characteristics.

Condor DAGMan Stork (Kosar *et.al*, 2004) was developed as a a batch scheduler specialized in data placement and data movement which understands the semantics and characteristics of data placement tasks and implements techniques specific to queuing, scheduling, and optimization of these type of tasks. Stork acts like an I/O control system (IOCS) between the user applications and the underlying protocols and data storage servers. The users can easily add support for their favorite storage system, data transport protocol, or middleware. Stork can interact with higher level planners and workflow managers, allowing the users to schedule both CPU resources and storage resources together.

Pegasus (Deelman *et.al.*, 2005) enables scientists to construct workflows in abstract terms without worrying about the details of the underlying cyberinfrastructure or the particulars of the lowlevel specifications required by the cyberinfrastructure middleware. As part of the mapping, Pegasus automatically manages data generated during workflow execution by staging them out to user-specified locations, by registering them in data catalogs, and by capturing their provenance information. Pegasus dynamically discovers the available resources and their characteristics, and queries for the location of the data (potentially replicated in the environment).

Swift uses for data management Virtual Data System (Foster *et. al.*, 2003) - a framework which provides a suite of components and services for data-intensive sciences that enables scientists to systematically and efficiently describe, discover, and share large scale data and computation resources. A common data model is used to describe data types and representations, and the recipes for derivations of data are specified in a declarative manner. Requests for data products can be transparently mapped into computation and/or data access operations across multiple Grid computing and storage locations.

## 4. FAULT TOLERANCE APPROACHES

Due to the heterogeneous and distributed nature of Grid systems, faults inevitably happen. The reasons for faults in a Grid environment are manifold: the geographically widespread nature encompassing multiple autonomous administrative domains, variations in the configuration of the different systems, overstrained resources that may stop responding or show unpredictable behavior, faults in the network infrastructure that connects the systems, hardware failures and systems running out of memory or disk space are just some of the possible sources of faults.

The taxonomy by (Yu *et.al.*, 2005) introduces a general view of existing workflow managing solutions. A part of it focuses on fault tolerance, where they use a task and workflow-level division. These categories can be further extended, keeping the scope on fault tolerance, detailing description and comparison of their properties. (Hwang *et. al.*, 2003) proposes a multi-layered approach for fault tolerance in workflows. They also segment the techniques into task-level and workflow-level techniques. The former tries to hide faults that happen during the execution of single tasks at the workflow-level, while the latter manipulates the structure of the workflow to deal with faults dynamically.

Besides these, several layers cand be identified where detection as well as recovery and prevention may exist:

- **Hardware level** - lowest loevel, machine crashes and network connectivity errors can happen.

- **Operating Systems level** - tasks may run out of memory or disk space, or exceed CPU time limits or disk quota. Other faults like network congestion or file nonexistence can also happen.

- **Middleware level** - non-responding services probably caused by too many concurrent requests. Authentication, file staging or job submission failures can happen, and submitted jobs could hang in local queues, or even be lost before reaching the local resource manager.

- **Tasks level** - job-related faults can happen, like deadlock, livelock, memory leak, uncaught exceptions, missing shared libraries or job crashes, even incorrect output results could be produced.

- **Workflow level** - failures can occur in data movement or infinite loops in dynamic workflows. Incorrect or not available input data could also produce faults.

- **User level** - user-definable exceptions and assertions which can cause errors.

Regarding fault prevention and recovery from faults, we can distinguish among three abstraction levels. The treatment mechanisms can act at Task, Workflow or User-level.

- At the Task level, recovery is used when a failed job is restarted on the same remote resource or resubmitted to another one. Generally it is simple to implement this technique; upon detecting a failure, the task is rescheduled to either the same or to another resource for another try. Resubmission can cause significant overheads if the following tasks have to wait for the completion of the failed task. Saving checkpoint and restarting later or even migrating jobs can be a good prevention and recovery mechanism. This technique stores all the intermediate data of a task that is needed to restore the task to the current state. This allows for migration of a task to another system in case of failure: it can resume execution from the last checkpoint, unlike simple resubmissions, where jobs should be started over from the beginning. Task replication can prevent resource failures, while alternate task creation can recover from internal task failures (in this case another task implementation is executed). On failures of the task manager itself, recovery means restarting the service or choosing another one. Finally resource reliability measurements can also prevent job execution faults.

- At Workflow level, redundancy, data and workflow replication can prevent faults. Redundancy, sometimes called replication in related work, executes one task concurrently on several resources, assuming that one of the tasks will finish without a failure. It can cause overhead by occupying more resources than necessary, but guarantees failure-free execution as long as at least one task does not fail. Light- and Heavy-weight checkpointing can also be used for both prevention and recovery. Generally this technique can be used to save an intermediate state of a whole workflow for a restart at a later point in time. Light-weight checkpointing saves only the current location of the intermediate data, not the data itself. It is fast, but restarting can only work as long as the intermediate data is available at its original location. Heavy-weight checkpointing saves all the intermediate data to a place, where it can be kept as long as it is needed.

- At the User level, user-defined exceptions can be taken into account to validate proper execution. The questionnaire also contained two tables for this section: the first is used to tell whether the listed mechanism is supported or not, the second is for naming the service that handles the faults.

Studying several workflow managers' (Pegasus, Askalon, P-GRADE, Triana) fault tolerance, prevention and recovery capabilities at all these levels (Plankensteiner *et.al.*, 2007) observed that Hardware-level faults (Machine crashed/down, Network down) can generally be successfully detected by current workflow systems. When it comes to the other categories, the situation is quite different. On the OS-level, only 37% of the faults (Disk quota exceeded, Out of memory, Out of disk space, File not found, Network congestion, CPU time limit exceeded) are currently detected on average.

Detection of the faults on middleware level (Authentication failed, Job submission failed, Job hanging in the queue of the local resource manager, Job lost before reaching the local resource manager, Too many concurrent requests, Service not reachable/not responding, File staging failure) is more common, an average of 62.8% of these faults can be detected by current Grid workflow systems, which is almost the same within Workflow-level faults (Infinite loop, Input data not available, Input error, Data movement failed) with 62.5%. The worst fault detection can be seen on the Task-level (Memory leak, Uncaught exception, Deadlock/Livelock, Incorrect output data, Missing shared libraries, Job crashes) and User-level (User-definable exceptions, User-definable assertions), where only 30% (task-level) and 25% (user-level) of the faults are detected on average.

While the workflow paradigm, emerged from the field of business processes, has been proven to be the most successful paradigm for creating scientific applications for execution also on Grid infrastructures, most of the current Grid workflow management systems still cannot deliver the quality, robustness and reliability that are needed for widespread acceptance as tools used on a day-to-day basis for scientists from a multitude of scientific fields.

Therefore our approach aims a configurable mechanism for fault tolerance: users can select whether they want to re-execute a job, to save partial results, to replicate job execution, to create a "compensation" mechanism (ex: exception handling). This approach is determined by the different semantics of workflows, in some case re-execution being a good solution but in other cases this can cause side effects.

## 5. THE ARCHITECTURE AND FUNCTIONALITY OF THE WORKFLOW MANAGEMENT PLATFORM

As we have shown in the previous sections, although several open source workflow engines are available for executing scientific applications in distributed environments, most of them lack important features concerning fault tolerance, abstract workflows, data handling and user interface. We note however that some of the existing engines are based on highly expressive languages and provide advanced process management, transaction handling, database persistence and other mechanisms. As a consequence, we chose the solution is of starting from an open source workflow engine and building additional modules to satisfy our requirements.

The workflow language we propose for the platform is WS-BPEL, which is a widely adopted standard in industry and, more recently, in academic environments. In what concerns the base workflow engine, we propose ActiveBPEL, which is the most frequently used open source BPEL engine and has been integrated in several research projects; some of the projects, like the one presented in (Subramaninan *et.al.*, 2008), also augmented WS-BPEL with additional modules. We briefly describe as follows the ActiveBPEL architecture and the extensions we intend to implement for our project.

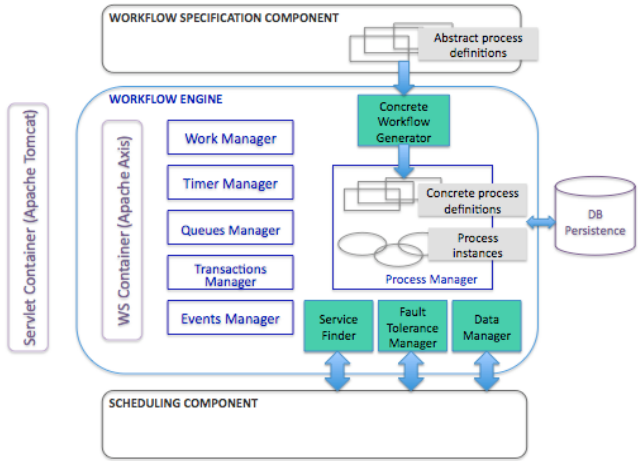

Fig. 1. ActiveBPEL-based workflow engine architecture. The new modules with which we propose to extend ActiveBPEL are depicted in green.

ActiveBPEL runs on top of the Apache Tomcat servlet container, and uses an embedded version of Apache Axis for message communications. Figure 1 presents the main components of ActiveBPEL (in blue) and our proposed extensions (in green). Among the services used in ActiveBPEL for handling processes, which are named Managers, the most important one is the Process Manager. The Process Manager oversees the instantiation and execution of processes and activities. When a process is deployed, the engine analyzes the BPEL sources and generates an internal representation of the process; then, when the user requires the execution of the process, a new instance is created by the Process Manager. The Process Manager is also responsible with instantiating activities and associating them with states (inactive, executing, finished, faulted etc.) during their life cycle. The Queue Manager handles incoming messages and events addressed to the process activities, by building a queue with the activities that are waiting for messages. The Work Manager schedules asynchronous operation, based on "work objects" which are a specialized alternative to threads. We also mention the Time Manager, which provides support for timed operations (like suspending or waiting), and the Transaction Manager, which implements methods for working with transactions.

We propose to introduce the following new components in the ActiveBPEL engine:

- **Concrete Workflow Generator**, which will transform abstract workflows into concrete workflows

- **Service Finder**, which will map service port types with sets of corresponding available services

- **Fault Tolerance Manager**, which will apply the policies specified by the user for handling faults

- **Data Manager**, which will implement efficient data handling mechanisms.

*5.1. Support for Abstract Workflow Specifications*

The Concrete Workflow Generator will have as inputs the abstract workflows specified by the users with the aid of the Workflow Specification component; its role will be to perform the mapping between abstract functional components and web service port types or executable programs. As we focus on scientific applications we are concerned about the particular design of these workflows which typically requires the involvement of at least two domains: one from the scientific field of interest (e.g. high energy physics, molecular biology) and another from computer science - understanding the process of composing the workflow and encoding the derivation in a format that the engine can execute. Because these domain have distinct terminology to describe workflow elements, including requirements, clear specification and effective mapping are a challenge.

Our approach for abstract workflow specification uses ontologies, as they are used to describe knowledge about a domain such that its representation can be interpreted and reasened about by a computer. We use ontologies first as an explicit specification of abstract concepts and later to support the composition and matching of services.

While domain expert's workflow descriptions are more often abstract, our engine needs a concrete specification of an executable workflow. We therefore opted for the use of BPEL for Semantic Web Services (BPEL4SWS) (Nitzsche *et. al.*, 2007) as a means to increase productivity during the design of workflows in support of scientific applications. BPEL4SWS introduces the desired level of abstraction for modeling workflows that is consistent with the target domain. It is thereby used by our Concrete Workflow Generator component to automatically generate executable workflows, that is, workflow implementations.

In our proposed architecture, BPEL4SWS uses Semantic Web Service Frameworks to define a communication channel between two partner services instead of using the partner link which is based on WSDL 1.1. It enables describing activity implementations in a much more flexible manner based on ontological descriptions of service requesters and providers. The specification introduces an extension to BPEL to enable describing interaction using semantic Web service Frameworks instead of using WSDL 1.1. Semantic Web services (SWS) can be considered an integration layer on top of Web services; they use ontologies as data model and they have a rich conceptual model. There are efforts towards standardizing this conceptual model within the Reference Ontology for Semantic Service Oriented Architectures (RO4SSOA). In addition to the SWS based interaction, BPEL4SWS makes use of annotated data types to enhance data handling by means of ontological mediators and uses ontological reasoning to evaluate conditions. Our Concrete

Workflow Generator component receives a BPEL4SWS specification as input and translates it into WS-BPEL, used by the ActiveBPEL engine to execute the workflow.

*5.2. Dynamic Web Services Composition*

The Service Finder will contact the scheduling component in order to discover web services (ports) that correspond to the port types specified in the workflow, using a find-and-bind approach similar with the one presented in (Miles *et. al.*, 2007a). We aim at transparently adapting existing composite services to encapsulate autonomic behavior (Kephart *et. al.*, 2003). That is, making composite services adaptable to changes in their execution environment (e.g., failure in a partner Web service). Although this is a major concern in the field of composite services, it is often not addressed in the specification of composition languages.

Our Service Finder component maps the abstract nodes onto matching services iteratively during the processing of the workflow. Each time the workflow engine reaches a transition related to an abstract (non-executable) operation, it calls a special workflow refinement service. This service refines the workflow description by searching for matching service candidates, which fulfill the requirements defined by the profile of the abstract nodes. The decision of whether a service matches the requirements is done by rules that depend on several properties, such as functionality (e.g., service produces certain class of output data or side effect), performance (e.g., operation should complete within 1h), or reliability (e.g., only services which have been operational during the last 72h should be taken into account). If it is possible to find matching service candidates, the refinement service attaches a list of the corresponding interface descriptions URLs (e.g. wsdl URLs) to the abstract transition. Next, the binding consists of the selection of one service instance out of the list of available service candidates at runtime. In order to optimize this dynamic selection, the system uses input from the scheduling component of the platform, which takes into account the recorded as well as the current monitoring information about the services and the Grid infrastructure.

*5.3. Fault Tolerance*

The Fault Tolerance Manager's role is to attempt the recovery after an activity failure, by applying one of the available policies: re-try the activity, find an alternative service to invoke, save the partial results; activity replication is another approach that the user will be able to choose. A significant drawback of existing workflow systems is their poor support for exception handling. Our component aims at identifying the specific error conditions which occurred and taking consequent actions. Hence, the Fault Tolerance Manager distinguishes and reports the exceptions to which is confronted: failures of invoked applications, communication failures, lack of response from a user, missed deadlines, and unexpected behavior of applications. This is achieved by subscribing to the Queue Manager of the ActiveBPEL engine and inspecting all error related messages.

The usual failure-handling procedure in most systems is to stop process execution and report the failure to an administrator. However, as workflow applications become larger and more complex, manual failure resolution becomes less and less feasible because of the demand for human resources, with their high cost and slow answer time. Clearly, we need automatic exception handling, especially for scalable systems. Therefore, our approach automatically applies the hierarchy of policies defined by users. Thus, the Actions module of the Fault Manager component is able to take action when some configurable condition is met. This way, when a given threshold is reached, an alert e-mail can be sent, or a program can be run, or an instant message can be issued. Actions represent the first step towards the automation of the management decisions in scientific workflows.

*5.4. Data Management Functionalities*

For the efficient management of the workflow data, we propose to introduce the Data Manager component, which will contact the underlying middleware in order to find mappings between logical and physical file names, and will generate metadata that will allow making associations between files and the applications that produced or modified them. We extend ActiveBPEL with disk usage optimization techniques by implementing the algorithm presented in (Ramakrishnan *et. al.*, 2007). Hence, we minimize the disk space footprint of scientific workflows by removing data as soon as it is no longer needed and scheduling the workflow tasks by first taking into account the data requirements of the workflow and the data space availability at the resources.

## 6. CONCLUSIONS AND FUTURE WORK

As we have shown above, from studying the existing workflow languages and platforms we have concluded that most of the current platforms do not provide a complete coverage for aspects like abstract workflow specification, data management, fault tolerance and interoperability with multiple middleware systems. Our goal is to develop a workflow platform that can offer all these functionalities, and we believe that the best approach for achieving this goal is to introduce an additional set of components to an existing open source workflow engine. We chose the ActiveBPEL engine due to the fact that it is based on the WSBPEL language, which has the advantages of standardization and high expressivity, and also due to its large community of users. The next steps in this project are to elaborate more detailed specifications for the proposed workflow engine components, to define their interface with the platform's lower and higher levels and then to start the implementation. Performance is also an important concern, so we intend to apply a benchmark based method for comparing our platform with other similar engines.